\documentclass[twocolumn,showpacs,amsfonts,aps,prc,nofootinbib,floatfix,%
superscriptaddress]{revtex4}

\usepackage{float}

\usepackage{amsmath}
\usepackage{bm}
\usepackage{graphicx}

\usepackage{epsfig}
\newcommand{\beq}{\begin{equation}}
\newcommand{\eeq}{\end{equation}}
\newcommand{\bea}{\vspace{0.25cm}\begin{eqnarray}}
\newcommand{\eea}{\end{eqnarray}}

\newcommand{\ro}{\mbox{{\boldmath
$\rho$}}}

\def\lsim{\mathrel{\rlap{\lower4pt\hbox{\hskip1pt$\sim$}}
    \raise1pt\hbox{$<$}}}         %less than or approx. symbol
\def\gsim{\mathrel{\rlap{\lower4pt\hbox{\hskip1pt$\sim$}}
    \raise1pt\hbox{$>$}}}         %greater than or approx. symbol

\begin{document}

%%%%%%%%%%%%%%%%%%%%%%%%Front Matter%%%%%%%%%%%%%%%%%%%%%%%%%%%%%%%%%%
%%%%%%%%%%%%%%%%%%%%%%%%%%%%%%%%%%%%%%%%%%%%%%%%%%%%%%%%%%%%%%%%%%%%%%
%\renewcommand{\thefootnote}{\fnsymbol{footnote}}

\title{\Large\bf
Jet quenching for hadron-tagged jets in 
$pA$  collisions
}

\date{\today}

\author{B.G. Zakharov}

\address{
L.D.~Landau Institute for Theoretical Physics,
        GSP-1, 117940,\\ Kosygina Str. 2, 117334 Moscow, Russia
}

\begin{abstract}
  We calculate the medium modification factor $I_{pA}$
for 5.02 TeV $p$+Pb collisions.
We use the Monte-Carlo Glauber model to
determine the parameters of the quark-gluon plasma fireball in $pA$
jet events.
Our calculations show that the jet quenching effect for $I_{pA}$
turns out to be rather small.
We have found that the theoretical $I_{pA}$ as a function
of the underlying event charged multiplicity density, within errors,
agrees with data from ALICE \cite{ALICE_Ipp_PLB} for 5.02 TeV $p$+Pb
collisions.
However, the experimental errors are too large to
draw a firm conclusion on the possible presence of jet quenching.\\

\end{abstract}
%
%\pacs{12.38.Mh, 24.85.+p}

\maketitle
%-------------------------------------------------------------
{\bf Introduction}.
  The transverse flow effects and the strong suppression of high-$p_T$
hadron spectra (jet quenching) observed in $AA$ collisions at RHIC and the LHC
are strong evidence for formation of the quark-gluon plasma (QGP)
(for reviews see, e.g., \cite{hydro2,hydro3,JQ_rev}).
The results of hydrodynamic modeling of the flow effects \cite{hydro2,hydro3}
and of the jet quenching analyses of the
heavy ion data on the nuclear modification factor $R_{AA}$
(see, e.g., \cite{CUJET3,Armesto_LHC,Z_hl})
are consistent with the onset of the collective/hydrodynamic behavior of
the QCD matter at the proper time $\tau_0\sim 0.5-1$ fm.
Although the dynamics of the QCD matter in $AA$ collisions
was under active investigation in the last years
for the weak and strong coupling scenarios (for reviews see, e.g.,
\cite{Teaney1,Strickland1,chesler1}),
the mechanism of the early
hydrodynamization of the QGP is not well understood yet.
The observation of the QGP formation in $AA$ collisions
has sparked new interest in the idea
of the QGP formation for small systems \cite{Shuryak_QGP}.
The analysis of \cite{Chesler2} in the scenario of the strongly coupled
QGP shows that even the smallest droplet of the QCD matter
produced in $pp/pA$ collisions can be described within hydrodynamics.
In \cite{Spalinski} it was argued that for $pp/pA$ collisions
the lower bound for applicability
of hydrodynamical description is $dN_{ch}/d\eta \sim 3$.
This is close to the estimate of \cite{RMK1609}. 

Experimentally, the formation of a mini QGP in $pp/pA$ collisions is
supported by 
the observation of the ridge effect
\cite{CMS_ridge,ATLAS_mbias,ATLAS_ridgepA,ALICE_ridgepA} in $pp/pA$
collisions at the LHC energies and by the steep growth  of
the midrapidity strange particle production
at charged multiplicity $dN_{ch}/d\eta\gsim 5$ \cite{ALICE_strange}.
The earlier analysis \cite{Camp1} of $\langle p_{T}\rangle$ as a function of
multiplicity, employing Van Hove's arguments \cite{VH}
also supports the onset of QGP regime at such charged multiplicity density.
There were suggested alternative non-hydrodynamical explanations of the
ridge effect in $pp/pA$ collisions \cite{Kovner_ridge1,Dumitru_ridge1,ridge_CGC}
due to the initial state parton effects. However, these models do not
explain the anomalous variation with the charged multiplicity density of
the midrapidity strange particle production and of $\langle p_{T}\rangle$.

Besides the generation of the flow effects, the QGP formation in $pp/pA$
collisions should lead to jet modification due to parton
energy loss in the QGP fireball.
It is important that 
the typical charged multiplicity of soft (underlying event (UE)) hadrons in
jet events is bigger than the ordinary minimum bias
multiplicity by a factor of $\sim 2-2.5$ \cite{Field}.
For the LHC energies the typical midrapidity charged
multiplicity in $pp$ jet events $dN_{ch}/d\eta\sim 10-15$.
One of the possible experimental methods to probe jet quenching in
  the small size QGP produced in $pp/pA$ collisions
  is investigation of the UE multiplicity dependence of the
  jet fragmentation functions (FFs) for photon/hadron
  tagged jets \cite{Z_pp_PRL} described by the
  modification factors $I_{pp/pA}$.  Formally, $I_{pp,pA}$ can
  be defined as the ratio of the per-trigger particle ($h^t$) yield
  of the associated  hadron ($h^a$) production,
  $Y_{pp,pA}$, to the yield  for $pp$ collisions calculated
  without the medium effects, $Y_{pp}^0$. However, $Y_{pp}^0$ is
  unobservable quantity.
  For this reason, it is convenient to characterize the medium effects
  in $pc$ collisions in terms
of the UE multiplicity dependence of the ratio of the experimental
yields $Y_{pc}$ and the average yield for $pp$ collisions
$\langle Y_{pp}\rangle$ 
\beq
  \frac{Y_{pc}(\{p_T\},\{y\})}
{\langle Y_{pp}(\{p_T\},\{y\}) \rangle}\,,
\label{eq:10}
\eeq
where $\{p_T\}=(p_T^{a},p_T^{t})$ and $\{y\}=(y^{a},y^{t})$ are the sets of
the transverse momenta and rapidities
of the trigger particle and the associated hadron, and
$\langle... \rangle$ means averaging over the UE multiplicity.
In terms of the modification factors $I_{pc}$ (defined via the unobservable
yield $Y_{pp}^0$)
the ratio (\ref{eq:10}) should be equal to the ratio
$I_{pc}/\langle I_{pp}\rangle$.
  Recently,  the ALICE collaboration \cite{ALICE_Ipp,ALICE_Ipp_PLB}
measured the UE multiplicity dependence
of the ratio (\ref{eq:10}) for the hadron
  tagged jets in $pp$ and $p$+Pb collisions at $\sqrt{s}=5.02$ TeV.
The ALICE \cite{ALICE_Ipp_PLB} measurement has been performed 
for the hadron momenta
$8<p_T^t<15$ GeV, $4<p_T^a<6$ GeV, and the UE activity has been
characterized by
the charged multiplicity $N_{ch}^{T}$ in the transverse 
kinematical region $\pi/3\leq |\phi|\leq 2\pi/3$,
$|\eta|<0.8$, and $p_T>0.5$ GeV. As compared to the UE charged
multiplicity density $dN_{ch}^{ue}/d\eta$, defined in the whole $\phi$ and $p_T$
regions for the pseudorapidity window $|\eta|<0.5$, $N_{ch}^{T}$
of \cite{ALICE_Ipp,ALICE_Ipp_PLB} is smaller by a factor of $\approx 4.4$.
For $pp$ collisions, in \cite{Z_Ipp} it was found that $I_{pp}$ decreases 
by about 7-10\% with increase of the UE activity in the range
$5\lsim dN_{ch}/d\eta\lsim 20$ for the jet quenching calculated
within the light-cone path integral formalism \cite{LCPI1,LCPI2004,RAA04,RAA08}
for the induced gluon emission.
The results of \cite{ALICE_Ipp} for $I_{pp}/\langle I_{pp}\rangle $ agree qualitatively with calculations
of \cite{Z_Ipp}. 
%%%%%%%%%%%%%%%%%%%%%%%%%%%%%%%%%%%%%%%%%%%%%%%%%%%%%%%%%%%%%%%%
It would be interesting to perform calculation of $I_{pA}$ and comparison
with data from \cite{ALICE_Ipp_PLB} as well.
The data of \cite{ALICE_Ipp_PLB} for $Y_{pA}$ also show a tendency of
some decrease of $Y_{pA}$ with increasing the UE charged multiplicity.
But the effect seems to be somewhat smaller, at least visually,
than that observed for $pp$ collisions.
However, one should bear in mind that in the case of $pA$ collisions the observed UE charged multiplicity density
may contain a considerable fraction of hadrons that come
from interaction of the projectile with peripheral nucleons
without the formation of the collective QCD matter. 
Because interaction with the peripheral nucleons may produce low
density/entropy parton system,
for which the energy/entropy density is not large enough to form
the QGP. Thus, one can expect that the fireball of the QCD matter
in $pA$ collisions should have the core-corona structure
(discussed previously for $AA$ collisions \cite{corona1}).
The effect of hadrons from the corona region on jet quenching should be small
since these hadrons should be in the free streaming regime.
For this reason, for
the jet quenching calculation
of the variation of $I_{pA}$ with the observed UE charged multiplicity
$dN_{ch}^{ue}/d\eta$ one needs a formalism for accounting for
the difference between the observed $dN_{ch}^{ue}/d\eta$ and
the charged multiplicity related to formation of the QGP
fireball (which we denote by $dN_{ch}^{f}/d\eta$).
In the present paper we perform such jet quenching calculations
of $I_{pA}$ for conditions of the ALICE experiment \cite{ALICE_Ipp}
using the Monte-Carlo (MC) Glauber model for calculation of the QGP
fireball parameters as a function of the total UE charged multiplicity
density. The parameters of the QGP fireball depends on the
free parameters of the MC Glauber model. However, our results demonstrate
that predictions for $I_{pA}$ turn out to be quite stable to
the theoretical uncertainties of the MC Glauber scheme.\\

{\bf Theoretical framework for calculation of $I_{pA}$.}
  Similarly to $AA$ collisions \cite{Wang_di-h,PHENIX_di-h},
  the per-trigger yield $Y_{pA}$ for production of the trigger hadron
$h^t$ and the associated hadron $h^a$
can be written via the di-hadron (back-to-back) and one-hadron
inclusive $pN$ cross sections as
\beq
Y_{pA}(\{p_T\},\{y\})=\left\langle\frac{d^4\sigma_{pN}}{dp_T^adp_T^tdy^ady^t}
\Big/\frac{d^2\sigma_{pN}}{dp_T^tdy^t}\right\rangle\,,
\label{eq:20}
\eeq
where $\langle \dots\rangle$ means averaging over the geometry of 
hard $pA$ collision. In particular, this averaging includes
averaging over the impact parameter for $pA$ collision. However,
it is natural to assume that the medium effects for $pA$ jet production
are sensitive only to the parameters of the QGP fireball,
and not to the impact parameter itself\footnote{In principle, the impact
  parameter dependence of $Y_{pA}$ can come from the nuclear modification
  of the nuclear PDFs. However, since $Y_{pA}$ (\ref{eq:20}) is expressed
  through the ratio of the hard cross sections, the effect of
  of the nuclear PDFs turns out to be practically negligible}.
For this reason, for a given UE charged multiplicity $N_{ch}^{ue}$
(i.e., for a given parameters of the QGP fireball) 
in (\ref{eq:20}) $\langle \dots\rangle$ is reduced to averaging
over the geometry of the jet production point  and of the jet trajectories
in the QGP fireball.
As in our previous calculations of $R_{pp}$ \cite{Z_hl} and $I_{pp}$
\cite{Z_Ipp},
we calculate $I_{pA}$ in the model of an effective azimuthally symmetric
QGP fireball.
We evaluate the distribution in the transverse plane of
the jet production points in $pN$ collisions for Gaussian
quark density (assuming that gluons have the same transverse distribution
as quarks).

We perform calculations of the hard medium modified $pN$ cross
sections in the
  collinear approximation, neglecting the
internal parton transverse momenta in the colliding nucleons.
The di-hadron and one-hadron cross sections can be written
as
\bea
\frac{d^4\sigma_{pN}}{dp_T^adp_T^tdy^ady^t}
=\int \frac{dz^t}{z^t}D_{h^t/i}^m(z^t,p_{Ti})\nonumber\\
\times D_{h^a/j}^{m}(z^a,p_{Tj})
\frac{d^3\sigma_{ij}}{p_{Ti}dp_{Ti}dy_idy_j}\,,
\label{eq:30}
\eea
\beq
\frac{d^2\sigma_{pN}}{dp_T^tdy^t}
=\int \frac{dz^t}{z^t}D_{h^t/i}^{m}(z^t,p_{Ti})
\frac{d^2\sigma_{i}}{dp_{Ti}dy_i}\,,
\label{eq:40}
\eeq
where
$\frac{d^3\sigma_{ij}}{dp_{Ti}dy_idy_j}$
is the two-parton cross section of $p+N\to i+j+X$ process for
$y_i=y^t$, $y_j=y^a$,
$p_{Ti}=p_{Tj}=p_{T}^t/z^t$ and $z^{a}=z^{t}p_T^a/p_T^t$, 
$\frac{d^2\sigma_{i}}{dp_{Ti}dy_i}$
is the one-parton cross section for $p+N\to i+X$ process,
$D_{h^t/i}^m$ and $D_{h^a/j}^m$
are the medium-modified FFs for transitions $i\to h^t$ and $j\to h^a$.

For calculation of
the medium-modified FFs $D_{h/i}^{m}(z,Q)$
we use the same method as in \cite{Z_hl}, and we therefore
  only briefly outline it.
We write $D_{h/i}^{m}$ 
as a $z$-convolution 
\beq
D_{h/i}^{m}(Q)\approx D_{h/j}(Q_{0})
\otimes D_{j/k}^{in}\otimes D_{k/i}^{DGLAP}(Q)\,,
\label{eq:50}
\eeq
where $D_{k/i}^{DGLAP}$ is the DGLAP FF for $i\to k$ transition,
$D_{j/k}^{in}$ is the in-medium $j\to k$ FF,
and  $D_{h^{a,t}/j}$ are the FFs for hadronization transitions of the
parton $j$ to hadrons $h^{a,t}$.
To compute the DGLAP FFs
$D_{k/i}^{DGLAP}$ we use the PYTHIA event generator \cite{PYTHIA}.
As in \cite{Z_hl}, we calculate the induced
gluon spectrum using the method of \cite{RAA04,RAA08}, and express the
in-medium FFs $D_{j/k}^{in}$
through the induced gluon spectrum in the approximation
of the independent gluon emission \cite{RAA_BDMS}.
We treat the collisional energy loss,
that is relatively small \cite{Z_coll},
as a perturbation to the radiative mechanism (see \cite{Z_hl}
for details).  
For the FFs $D_{h/j}$ we use the KKP
\cite{KKP} parametrization  with $Q_0=2$ GeV.

We calculate the gluon induced spectrum (needed
for calculation of $D_{j/k}^{in}$)
using the parametrization of running $\alpha_s$ 
in the form \cite{RAA20T,Z_hl}
(supported by the lattice results for the in-medium $\alpha_s$
\cite{Bazavov_al1}) 
\beq
\alpha_s(Q,T) = \begin{cases}
\dfrac{4\pi}{9\log(\frac{Q^2}{\Lambda_{QCD}^2})}  & \mbox{if } Q > Q_{fr}(T)\;,\\
\alpha_{s}^{fr}(T) & \mbox{if }  Q_{fr}(T)\ge Q \ge cQ_{fr}(T)\;, \\
\frac{Q\alpha_{s}^{fr}(T)}{cQ_{fr}(T)} & \mbox{if }  Q < cQ_{fr}(T)\;, \\
\end{cases}
\label{eq:60}
\eeq
with $c=0.8$,
$Q_{fr}(T)=\Lambda_{QCD}\exp\left\lbrace
{2\pi}/{9\alpha_{s}^{fr}(T)}\right\rbrace$ (we take $\Lambda_{QCD}=200$ MeV)
and $Q_{fr}=\kappa T$, where $\kappa$ is a free parameter.
We take $\kappa=2.55$. It has been obtained
for scenario with the QGP formation
in $pp$ collisions
by $\chi^2$ fitting of the LHC data on $R_{AA}$ for 2.76 and 5.02 TeV
Pb+Pb, and 5.44 TeV Xe+Xe collisions (see \cite{Z_hl} for details).  

  As in \cite{Z_hl}, we calculate $D_{j/k}^{in}$ for 
a fireball with a uniform entropy/density distribution in the transverse plane.
We assume the Bjorken 1+1D longitudinal \cite{Bjorken} expansion of 
the QGP fireball at the proper time $\tau>\tau_0=0.5$ fm. It corresponds
to the QGP entropy density $s=s_0(\tau_0/\tau)$. As in \cite{Z_hl},
for $\tau<\tau_0$ we take $s=s_0(\tau/\tau_0)$.
We write $s_{0}$ in terms of
the charged multiplicity density $dN_{ch}^{f}/d\eta$
as 
\beq
s_{0}=\frac{C}{\tau_{0}\pi R_{f}^{2}}\frac{dN_{ch}^{f}}{d\eta}\,,
\label{eq:70}
\eeq
where  $R_{f}$ is the fireball radius,
$C=dS/dy{\Big/}dN_{ch}/d\eta\approx 7.67$ is the entropy/multiplicity
ratio \cite{BM-entropy},
$dN_{ch}^{f}/d\eta$ is the part of the observed UE charged multiplicity
density related to
hadronization of the QGP fireball ($dN_{ch}^{f}/d\eta=F_{qgp}dN_{ch}^{ue}/d\eta$).

{\bf Calculation of the QGP fireball parameters}.
We calculate the fireball radius $R_f$ and the QGP fraction
of the UE multiplicity $F_{qgp}$ as functions of the observed
UE charged multiplicity density $dN_{ch}^{ue}/d\eta$ with
the help of the MC Glauber wounded nucleon model developed in
\cite{MCGL1,MCGL2}.
The MC Glauber  model of \cite{MCGL1,MCGL2} allows to perform calculations
accounting for the
nucleon meson cloud.
In the present analysis we use the version without the meson cloud.
The impact parameter  profile of the probability of inelastic
$NN$ interactions is taken in a Gaussian form 
$
P(\rho)=\exp\left(-\pi \rho^2/\sigma_{in}^{pp}\right)\,.
$
As in \cite{Z_xe}, for $\sigma_{in}^{NN}$ we use the NSD inelastic $pp$ cross
section $\sigma_{NSD}^{pp}=55.44$ mb.
For the $^{208}$Pb nucleus we use the Woods-Saxon nuclear distribution
$\rho_{A}(r)=\rho_0/[1+\exp((r-R_A)/a)]$
with the hard-core $NN$ repulsion. We take $R_A=6.407$ fm, $a=0.459$ fm, and
the hard-core radius $d=0.9$ fm \cite{GLISS3}.
In \cite{MCGL1,MCGL2,Z_xe} the MC Glauber simulations of nuclear collisions
have been performed including the the sources corresponding to the wounded
nucleons and to the binary collisions \cite{KN}.
Fitting of the LHC data on the centrality dependence of the midrapidity
$dN_{ch}(AA)/d\eta$ gives the binary collisions fraction parameter
$\alpha\sim 0.13$ \cite{MCGL2,Z_xe}. However,
the data on the midrapidity charged multiplicity density in
the minimum bias 5.02 TeV $p$+Pb collisions \cite{ALICE_pPb} agree better with
$\alpha\approx 0$.
In this case, the prediction of the Glauber wounded model for
the minimum bias midrapidity charged multiplicity
in $pA$ collisions reads
\beq
\frac{dN_{ch}^{mb}(pA)}{d\eta}=\frac{N_w^{mb}}{2}
\cdot\frac{dN_{ch}^{mb}(pp)}{d\eta}\,,
\label{eq:80}
\eeq
where $N_w^{mb}$ is the minimum bias number of wounded nucleons,
and $dN_{ch}^{mb}(pp)/d\eta$ is the minimum bias midrapidity charged multiplicity
density in $pp$ collisions. The MC Glauber simulation of minimum bias
5.02 TeV $p$+Pb collisions gives $N_w^{mb}\approx 6.87$. With this value and
$dN_{ch}^{mb}(pp)/d\eta\approx 5.34$\footnote{This value of
$dN_{ch}^{mb}(pp)/d\eta$  
  can be obtained \cite{Z_xe} with the help of the power 
law interpolation 
between the ALICE data \cite{ALICE_nch541} at $\sqrt{s}=2.76$ TeV
($dN_{ch}^{mb}(pp)/d\eta \approx 4.63$)  and 
at $\sqrt{s}=7$ TeV   
($dN_{ch}^{mb}(pp)/d\eta=5.74\pm 0.15$) for the charged multiplicity
in the NSD events.}, formula (\ref{eq:80}) gives
$dN_{ch}^{mb}(p\mbox{Pb})/d\eta\approx 18.3$ that agrees well with 
the ALICE measurement
$dN_{ch}^{mb}(pPb)/d\eta\approx 17.8$ \cite{ALICE_pPb}.
To generalize (\ref{eq:80}) to jet events 
one should account for the fact that one of the wounded nucleons in
the nucleus and the projectile proton participate in jet production.
It is reasonable to expect that the contribution of this pair $pN$
to the multiplicity density should be equal to the average
UE multiplicity density $dN_{ch}^{ue}(pp)/d\eta$ in $pp$ jet events.
Then, the generalization 
of (\ref{eq:80}) to the case of the minimum bias UE multiplicity density
in jet events can
be written as
\beq
\frac{dN_{ch}^{ue}(pA)}{d\eta}=\frac{dN_{ch}^{ue}(pp)}{d\eta}
+\frac{N_w^{ue}-2}{2}\cdot
\frac{dN_{ch}^{mb}(pp)}{d\eta}\,,
\label{eq:90}
\eeq
where $N_w^{ue}$ is the average number of the wounded nucleons
in jet events.
Our MC Glauber simulation gives $N_w^{ue}\approx 9.9$ for 5.02
TeV $p$+Pb collisions.
Then, using $dN_{ch}^{ue}(pp)/d\eta\approx 12.5$ for $5.02$ TeV $pp$ jet events
(obtained by interpolating the ATLAS data \cite{ATLAS_UE_Nch} at $\sqrt{s} =
0.9$ and $7$ TeV, assuming that
$dN_{ch}^{ue}(pp)/d\eta\propto s^{\delta}$), from the formula (\ref{eq:90})
we obtain $dN_{ch}^{ue}(p\mbox{Pb})/d\eta\approx 33.5$, which agrees well
with that obtained in the ALICE experiment \cite{ALICE_Ipp}
($\langle N_{ch}^T\rangle=7.77\pm 0.31$ that corresponds to
$dN_{ch}^{ue}(p\mbox{Pb})/d\eta=34.19\pm 1.36$).
Thus, one sees that the predictions of the MC Glauber model for the minimum bias
and the UE midrapidity charged multiplicities in $p$+Pb collisions
are in good agreement with the experimental data.

To perform the MC Glauber simulation of the geometry of the entropy
production in
$pA$ jet events, for each wounded nucleon we use the Gaussian
distribution of the entropy density  in the transverse coordinates  
\beq
\frac{dS_w^i(\ro-\ro_i)}{dy d\ro}=
\frac{C\exp[-(\ro-\ro_i)^2/\sigma^2]}{\pi\sigma^2}
\cdot\frac{dN^i_{ch}/d\eta}{2}\,,
\label{eq:100}
\eeq
where $i=1,2$ correspond to the nucleons participating in jet production,
and $i> 2$ to the ordinary wounded nucleons.
We treat $dN^i_{ch}/d\eta$ as random variables.
Below we use the short-hand notations $n_i$ for $dN^i_{ch}/d\eta$.
As in \cite{MCGL2,Z_xe}, we describe fluctuations
of $n_i$
by the widely used in the MC Glauber simulations Gamma distribution
\beq
G(n_i)=
\left(\frac{n_i}{\langle n_i\rangle}\right)^{\kappa_i-1}
\frac{\kappa_i^{\kappa_i}\exp\left[-n_i\kappa_i/\langle n_i\rangle\right]}
{\langle n_i\rangle \Gamma(\kappa_i)}\,.
\label{eq:110}
\eeq
The $\langle n_{i> 2}\rangle$ should be equal to the minimum
bias $pp$ charged multiplicity pseudorapidity density
$dN_{ch}^{mb}(pp)/d\eta$, i.e., for $5.02$ TeV $pp$ collisions
we have $\langle n_{i> 2}\rangle \approx 5.34$.
The value of the parameter
$\kappa_{i>2}$ has been adjusted to reproduce the experimental variance
of the charged $pp$ multiplicity in the unit pseudorapidity window
$|\eta|<0.5$. This gives
$\kappa_{i>2}\approx 0.56$ \cite{Z_xe}.
The value of $\langle n_{i=1,2}\rangle$ should be equal to
$dN_{ch}^{ue}(pp)/d\eta\approx 12.5$.
We adjusted the parameter
$\kappa_{i=1,2}$
to reproduce the
variance of the midrapidity $pp$ UE charged multiplicity density,
that has been evaluated using the CMS data \cite{CMS_UE7} on the UE
multiplicity distribution  for $7$ TeV $pp$
collisions (rescaled to $\sqrt{s}=5.02$ TeV assuming the KNO scaling).
This procedure gives
$\kappa_{i=1,2}\approx 1.1$

We used the following procedure to differentiate
the QGP fireball and corona transverse regions
in our MC Glauber simulation of the UE for jet production in $p$+Pb collisions.
We attribute to the QGP fireball the regions with the local ideal
gas temperature
(defined via the entropy density at $\tau=\tau_0$) larger
than $T_{min}\sim T_c$ (where $T_c\approx 160$ MeV is the deconfinement
temperature)\footnote{This criterion is similar to that used in
\cite{glasma_pp} within the IP-Glasma model (there the authors
attributed to the QGP regions with the energy density $\epsilon>
a\Lambda_{QCD}^4$ with $a\sim 1-10$).}.
We performed calculations for two versions with $T_{min}=160$
and $200$ MeV.
Using this procedure we obtained
the relative QGP contribution to the UE charged multiplicity
$dN_{ch}^{ue}/d\eta$ as functions of the observed UE charged
multiplicity density $dN_{ch}^{ue}/d\eta$.
We define the radius of the effective fireball (with the flat entropy
density), from the condition that the mean squared radius
of the QGP fireball coincides with that for the MC Glauber entropy distribution
in the region $T>T_{min}$. We also performed the calculations
for the radius of the effective fireball defined as $R_f=\sqrt{S_f/\pi}$,
where $S_f$ is the area of the region with $T>T_{min}$. It was found
that the results for these two methods are very close to each other. 

%%%%%%%%%%%%%%%%%%%%%%%%%%%%%%%%%%%%%%%%%%%%%%%%%%%%%%%%%%%%%%%%
\begin{figure} %[!h] %[!hbt]% [t]
\begin{center}
\includegraphics[height=5.5cm]{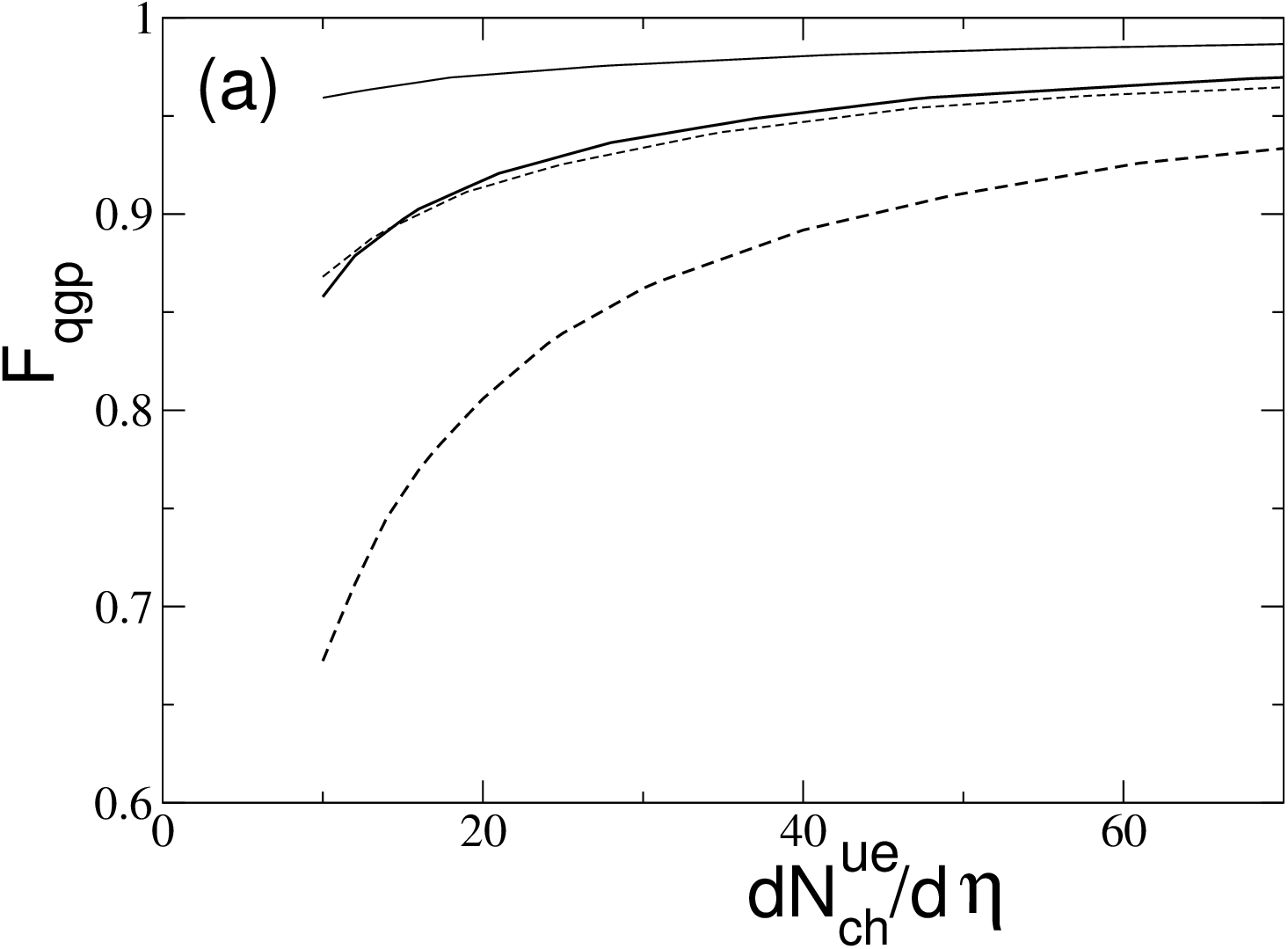}  
\includegraphics[height=5.5cm]{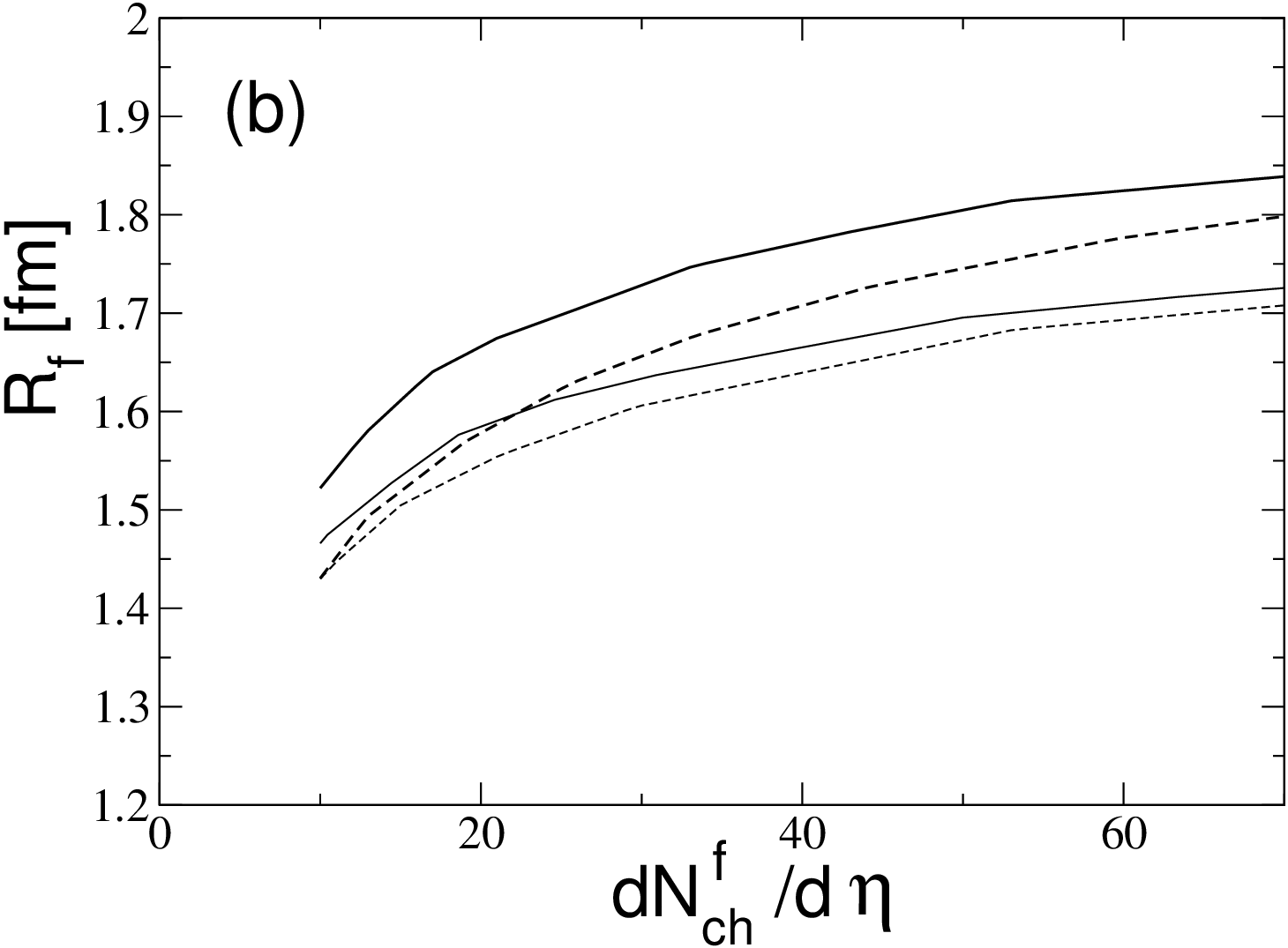}  
\end{center}
\caption[.]{The parameters of the QGP fireball obtained in the MC Glauber
  model simulations for $\sigma=0.7$ (thick lines) and
  $0.4$ fm (thin lines) for $T_{min}=160$ (solid) and $200$ (dashed) MeV.
(a) The relative contribution of the QGP fireball to the midrapidity UE charged
  multiplicity density $dN_{ch}^{ue}/d\eta$ vs $dN_{ch}^{ue}/d\eta$.
(b) The QGP fireball radius vs $dN_{ch}^{f}/d\eta$.  
 }
\end{figure}
\begin{figure} %[!h] %[!hbt]% [t]
\begin{center}
\includegraphics[height=5.5cm]{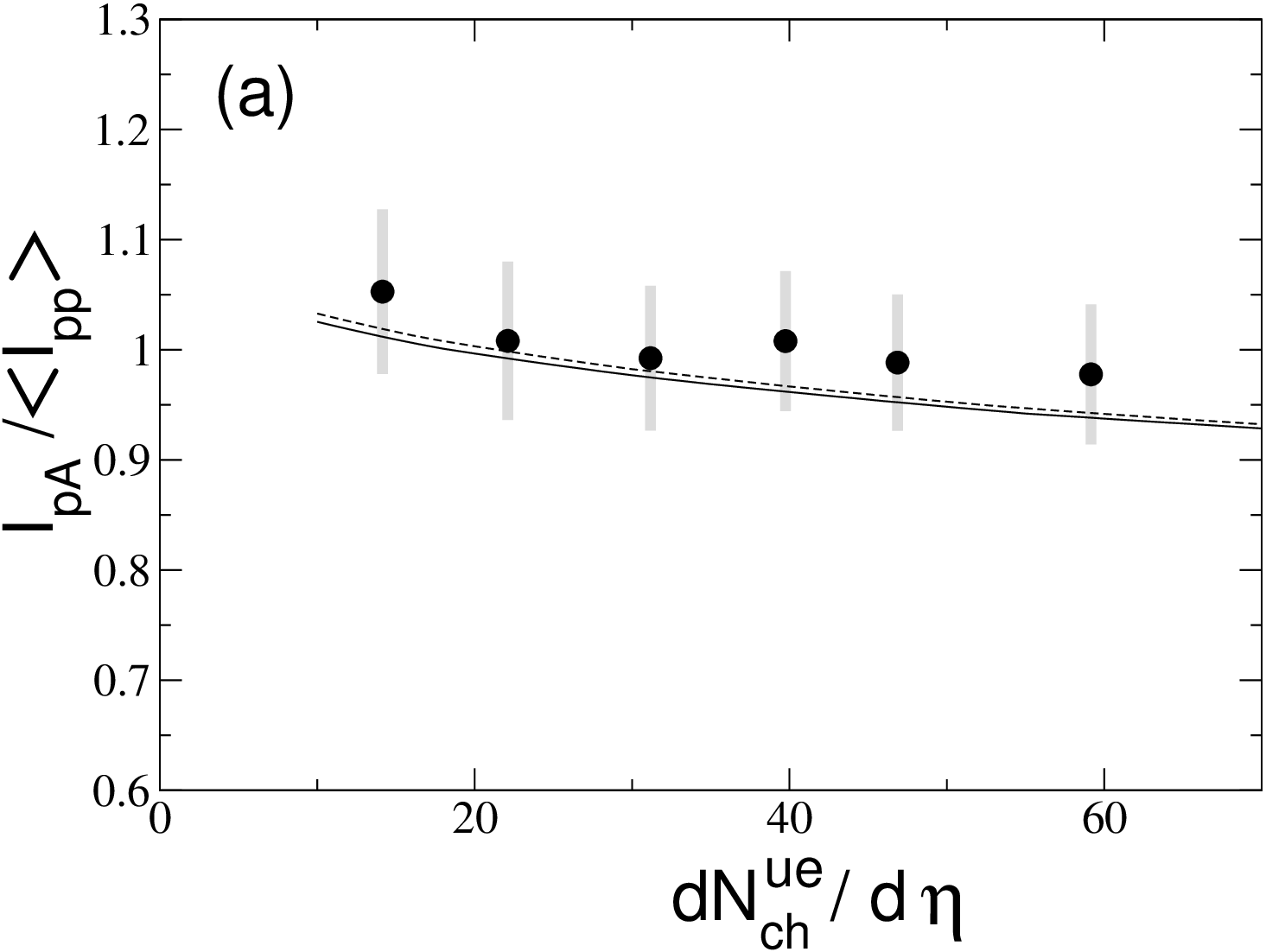}  
\includegraphics[height=5.5cm]{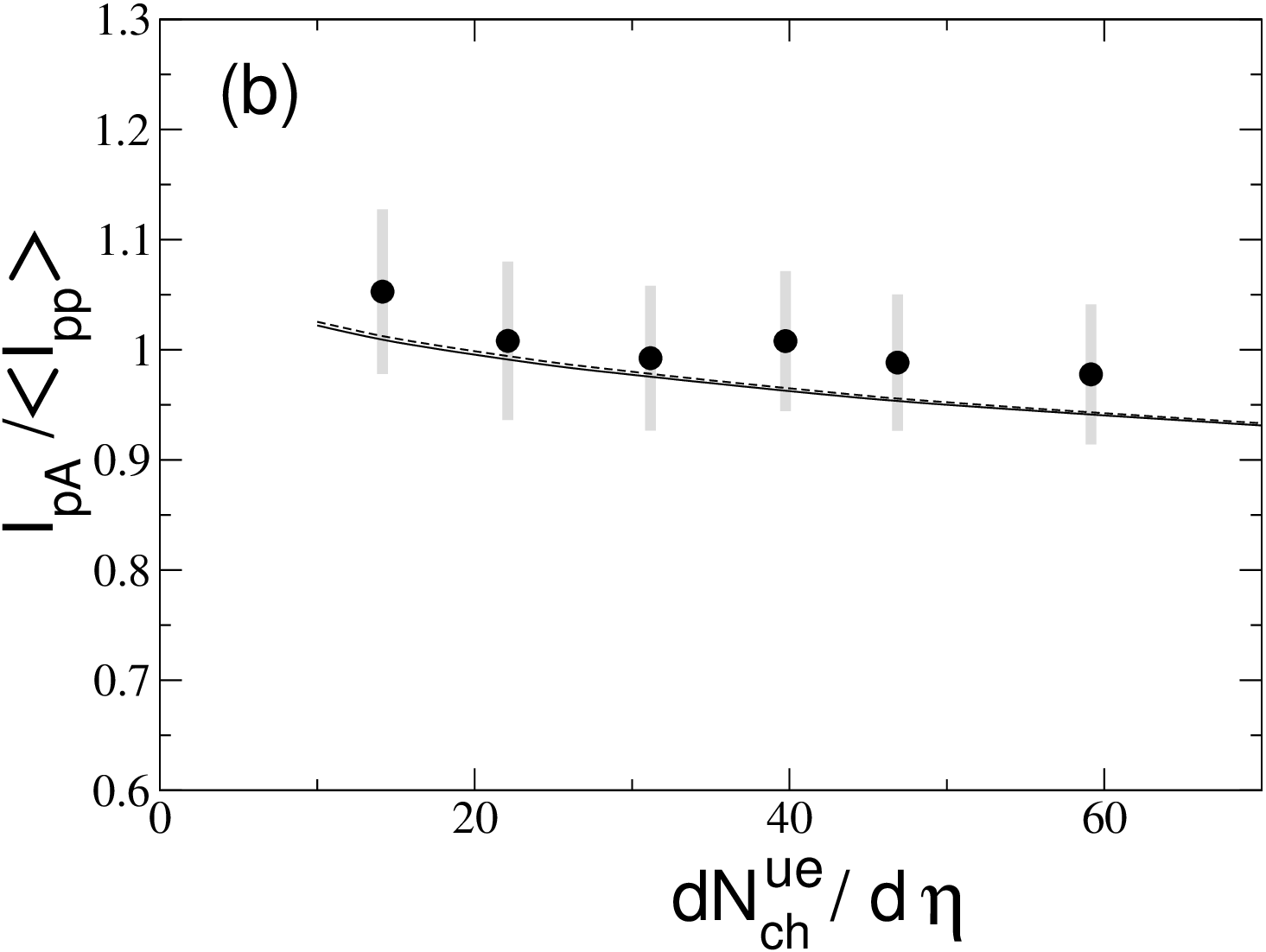}  
\end{center}
\caption[.]{The ratio $I_{pA}/\langle I_{pp}\rangle$ vs the midrapidity
  UE charged
  multiplicity density $dN_{ch}^{ue}/d\eta$ in $5.02$ $p$+Pb collisions
  obtained for $\sigma=0.7$ (a) and $0.4$ (b) fm for $T_{min}=160$ (solid)
  and $200$ (dashed) MeV. Data points are from \cite{ALICE_Ipp_PLB}
  (with the rescaled $N_{ch}^{T}$ variable used in \cite{ALICE_Ipp_PLB} by the
  factor 4.4 (see text for explanation).
 }
\end{figure}

{\bf Numerical results}.
To illustrate the sensitivity of our results to the choice of
the Gaussian width parameter $\sigma$
(entering Eq. (\ref{eq:100})) in the MC Glauber simulations of
the UE for jet production
in $pA$ collisions, we present the results for $\sigma=0.7$ and $0.4$ fm
(the latter value is often used in the MC Glauber model GLISSANDO
\cite{GLISS3}).
In Fig. 1a we plot the average relative QGP contribution
 to the UE charged multiplicity density $F_{qgp}$ as a function
of the UE charged midrapidity multiplicity density
$dN_{ch}^{ue}(pA)/d\eta$ obtained in the MC Glauber simulations
for $T_{min}=160$ and $200$ MeV.
As seen in Fig. 1a, $F_{qgp}$ becomes smaller at lower total UE charged
multiplicity, i.e., the corona effect is more important in the low
multiplicity jet events.
In Fig. 1b we present the results
for the effective fireball radius as a function
of $dN_{ch}^{f}/d\eta=F_{qgp}dN_{ch}^{ue}/d\eta$ for the same parameters as in
Fig. 1a. For the initial QGP temperature (for the ideal gas model)  we obtained 
$T_0\sim 210-220(330-350)$ MeV 
at $dN_{ch}^f/d\eta\sim 10(60)$.

The results for $I_{pA}/\langle I_{pp}\rangle$ as a function of the UE charged multiplicity density
are shown in Fig. 2 (we use $\langle I_{pp}\rangle\approx 0.94$ obtained
in \cite{Z_Ipp}).
The coordinates for the data points from \cite{ALICE_Ipp_PLB}, shown in Fig. 2,
have been rescaled
by the factor $r\approx 4.4$ from the variable $N_{ch}^T$ (corresponding
to the UE charged multiplicity in the transverse 
kinematical region $\pi/3\leq |\phi|\leq 2\pi/3$,
$|\eta|<0.8$, and $p_T>0.5$ GeV) to the UE charged
multiplicity $dN_{ch}^{ue}/d\eta$ defined in the whole $\phi$ and $p_T$
regions for the pseudorapidity window $|\eta|<0.5$.
The theoretical curves were obtained
for $\sigma=0.7$ (a) and $0.4$ (b) fm and $T_f=160$ (solid) and $200$ (dashed)
MeV.
Fig. 2 clearly shows that predictions for $I_{pA}$ turn out to
be rather insensitive to the parameter $\sigma$ in (\ref{eq:100}) used in the
MC Glauber model
and to the value of $T_{min}$ 
used
for evaluation of the size and the entropy of the QGP fireball.
We also performed calculations
for the MC Glauber scheme without the short range $NN$ correlations.
We have found that in this case the results are practically
the same as that shown in Figs. 1,~2. It is not surprising, since
for $p$+Pb collisions the typical 3D separation between the wounded nucleons
is larger than the $NN$ hard core radius by a factor of $2-3$.

As one can see from Fig. 2, the theoretical $I_{pA}$ decreases
by $\sim 8$\% from $dN_{ch}^{ue}/d\eta\sim 10$ to
$dN_{ch}^{ue}/d\eta\sim 60$.
The experimental data show approximately similar tendency.
The theoretical predictions are in reasonable agreement
with the data within the measured uncertainties.
However, from Fig. 2 one sees that the normalization of the
theoretical curves is smaller by $\sim 3-4$\% than that for
the ALICE data \cite{ALICE_Ipp_PLB}. This may be
due to the systematic errors of the experimental yield $Y_{pA}$, that are rather
large ($\sim 7$\%).
Physically, in the picture with jet modification
due to the final state interaction effects in the QGP fireball,
one can expect that in the region with
$dN_{ch}^{ue}(pA)/d\eta\sim \langle dN_{ch}^{ue}(pp)/d\eta\rangle\lsim 12-13$
(i.e., in the regime where the single $pN$ interaction should dominate)
there must be $Y_{pA}/\langle Y_{pp}\rangle\approx 1$, i.e.,
$I_{pA}/\langle I_{pp}\rangle\approx 1$.
But for the ALICE data \cite{ALICE_Ipp_PLB}
we have $Y_{pA}/\langle Y_{pp}\rangle\sim 1.05\pm 0.07$ at
$dN_{ch}^{ue}(pA)/d\eta\sim \langle dN_{ch}^{ue}(pp)/d\eta\rangle$.
In principle, the  double-parton
  scattering can lead to $Y_{pA}/\langle Y_{pp}\rangle>1$.
  However, the procedure of \cite{ALICE_Ipp_PLB} with calculating
  the ratio $Y_{pA}/\langle Y_{pp}\rangle$ using for the effective away side
  $pA$ and $pp$ yields the difference between the real measured away side
  yields and the measured yields in the transverse region, should automatically
  eliminate the possible background contribution from
  the double-parton scattering.
  It is worth noting that for the data of \cite{ALICE_Ipp_PLB}
  the factor $I_{pPb}$ at the maximal value of the $p$+Pb UE multiplicity
  ($N_{ch\text{max}}^T(pPb)\approx 13.4$) is larger than the
  factor $I_{PbPb}$ at
  the minimal Pb+Pb UE multiplicity ($N_{ch\text{min}}^T(PbPb)\approx 19.2$) by
  $\sim 10$\%, while, physically, one could expect $I_{pPb}/I_{PbPb}\approx 1$
  for $N_{ch}^T(pPb)\sim N_{ch}^T(PbPb)$ (since the UE multiplicity
  dependence of $I_{pPb}$ is weak, it is clear that the difference
  between $N_{ch\text{max}}^T(pPb)$ and $N_{ch\text{min}}^T(PbPb))$
  cannot lead to a considerable variation of $I_{pPb}/I_{PbPb}$).
  Note that in our model the factor $I_{pPb}$ at
  $dN_{ch}^{ue}(pPb)/d\eta\sim 85$
  (that corresponds to $N_{ch\text{min}}^T(PbPb)\approx 19.2$)
  agrees reasonably with the factor $I_{PbPb}$ of \cite{ALICE_Ipp_PLB}
  at the minimal $N_{ch}^T(PbPb)$.

{\bf Summary}.
We have calculated the medium modification factor $I_{pA}$
for hadron-tagged jets in 5.02 TeV $p$+Pb  collisions.
The medium-modified FFs have been evaluated within the
light-cone path integral approach to induced gluon emission.
We used parametrization of the running QCD coupling $\alpha_s(Q,T)$
which has a plateau around $Q\sim \kappa T$
(motivated by the lattice simulations \cite{Bazavov_al1}).
The value of $\kappa$ is fitted to
the LHC data on the nuclear modification factor $R_{AA}$
in $2.76$ and $5.02$ TeV Pb+Pb, and $5.44$ TeV Xe+Xe collisions.
To determine the size of the QGP fireball and the fraction
of the UE charged multiplicity density related to the formation
of the QGP fireball in $pA$ collisions we used the MC Glauber model.
We have found that the theoretical predictions for $I_{pA}$
are rather insensitivity to the theoretical uncertainties in the MC Glauber
modeling of the QGP formation.
Our calculations show that the jet quenching effect for $I_{pA}$
turns out to be rather small.
We have found that the theoretical $I_{pA}$ as a function
of the UE charged multiplicity density, within errors,
agrees with data from ALICE \cite{ALICE_Ipp_PLB} for 5.02 TeV $p$+Pb
collisions.
However, the experimental errors are too large to
draw a firm conclusion on the possible presence of jet quenching.

\begin{acknowledgments}
  I am grateful to S.~Tripathy
  for useful communication on some aspects of the
ALICE measurement of  $I_{pp,pA}$   
\cite{ALICE_Ipp}. 	
  This work is supported by Russian Science Foundation
  grant  No. 20-12-00200
  in association with Steklov Mathematical Institute.
\end{acknowledgments}

\end{document}